\def\arxiv{0}

\if\arxiv1
\documentclass{amsart}
\else
\documentclass{article}
\usepackage[T1]{fontenc} % add special characters (e.g., umlaute)
\usepackage[utf8]{inputenc} % set utf-8 as default input encoding
\fi
\usepackage{amsmath,cite,url}
\usepackage{graphicx}
\usepackage{color}

\usepackage{multirow}
\if\arxiv1
\usepackage{geometry}
\else
\usepackage{ismirarxiv}
\usepackage{lineno}
\fi

\usepackage{amsmath}
\usepackage{amssymb}
\usepackage{mathtools}
\usepackage{amsthm}

\theoremstyle{plain}

\theoremstyle{definition}

\theoremstyle{remark}

\newcommand{\R}{\mathbb R}

\newcommand{\up}{\textup}
\newcommand{\start}{\up{st}}
\newcommand{\nend}{\up{end}}
\newcommand{\pitch}{\up{pitch}}

\renewcommand{\tt}{\texttt}

\newcommand{\footnoteone}
{\footnote{Our system and video demo are available at \url{https://github.com/m-malandro/composers-assistant-REAPER}.}}

\newcommand{\footnotetwo}
{\footnote{\url{https://github.com/m-malandro/composers-assistant-REAPER}.}}

\title{Composer's Assistant: An Interactive Transformer for Multi-Track MIDI Infilling}

\if\arxiv1
\else
\oneauthor
  {Martin E.\ Malandro} {Sam Houston State University \\ \tt{malandro@shsu.edu}}
\def\authorname{M. E. Malandro}
\fi

\if\arxiv1
\usepackage{hyperref}
\else
\usepackage[bookmarks=false,pdfauthor={\authorname},pdfsubject={\papersubject},hidelinks]{hyperref}
\fi

\if\arxiv1
\author{Martin E.\ Malandro}
\title{Composer's Assistant: An Interactive Transformer for Multi-Track MIDI Infilling}
\address{Department of Mathematics and Statistics, Box 2206, Sam Houston State University, Huntsville, TX 77341-2206}
\email{malandro@shsu.edu}
\fi

\begin{document}

\if\arxiv1\else
\maketitle
\fi

\begin{abstract}
{
We introduce Composer’s Assistant, a system for interactive human-computer composition in the REAPER digital audio workstation.
We consider the task of multi-track MIDI infilling when arbitrary track-measures have been deleted from a contiguous slice of measures from a MIDI file, and we train a T5-like model to accomplish this task.
Composer's Assistant consists of this model together with scripts that enable interaction with the model in REAPER.
We conduct objective and subjective tests of our model. 
We release our complete system, consisting of source code, pretrained models, and REAPER scripts.
%To alleviate copyright concerns, our released models were trained only on MIDI files marked as being in the public domain, available under a CC0 license (or otherwise freely available to use without attribution), or available under a CC BY license, and MIDI files which we had permission from the authors to use for training.
Our models were trained only on permissively-licensed MIDI files.
}
\end{abstract}

\if\arxiv1
\twocolumn[\maketitle]
\fi

%\keywords{keywords HERE}

%\subjclass[2010]{05E30, 05B10}
%05E30   Association schemes, strongly regular graphs
%20M18 = inverse semigroups
%20C40 = Computational methods
%43A30 = Fourier and Fourier-Stieltjes transforms on nonabelian groups and on semigroups, etc.
%68W40 = Analysis of algorithms
%20M30 = Representation of semigroups; actions of semigroups on sets
%43A65 = Representations of groups, semigroups, etc. 
%62-07 = Data analysis

\section{Introduction}
\label{SecIntro}

\begin{figure}[t]
 \centerline{\framebox{
 \includegraphics[width=0.9\columnwidth, trim={0 0 0 202}, clip]{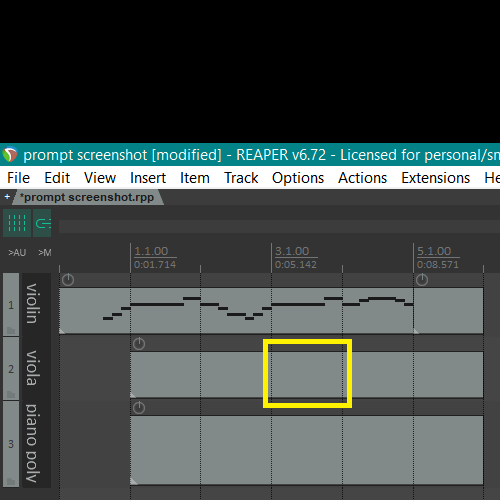}}}  % l b r t crop

 \centerline{\framebox{
 \includegraphics[width=0.9\columnwidth, trim={0 0 0 202}, clip]{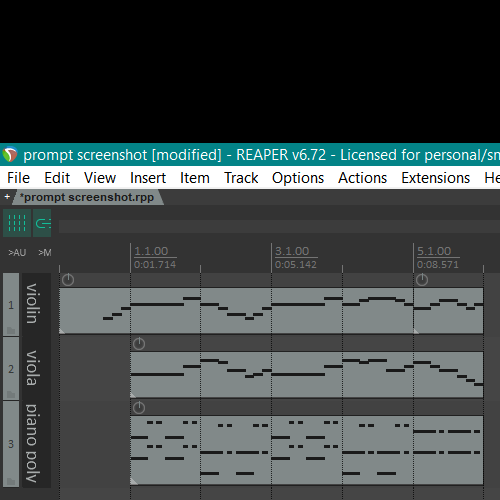}}}  % l b r t crop

 \caption{A prompt in REAPER, followed by a model output. Vertical lines separate measures. Users place empty MIDI items in REAPER to tell the model in which measures to write notes, and track names to tell the model what instrument is on each track. A track-measure in the prompt is boxed. Our model writes at least one note into every track-measure in every empty MIDI item in the prompt.}
 \label{FigPrompt}
\end{figure}

Many generative models for music exist. For instance, MuseNet \cite{musenet} and SymphonyNet \cite{symphonynet} can generate or continue a piece of music, and Music Transformer \cite{musictransformer} can continue a piano performance or harmonize a piano melody.
%, and MusicVAE \cite{musicvae} can generate up to 16-bar trios. 
When we tried using these tools as compositional aides, however, we quickly ran into limitations. For instance, while Music Transformer is capable of harmonizing a given melody, it does not offer the ability to keep part of the harmonization and regenerate the other part. MuseNet and SymphonyNet can generate a continuation of a user's prompt, but do not allow the user to regenerate individual instruments or measures within the continuation while keeping the rest of the continuation intact. 

DeepBach \cite{deepbach} can perform infilling on Bach-like chorales in a window specified by the user. Motivated by the idea of extending the DeepBach user experience to more styles, arbitrary collections of instruments, and arbitrary infilling target locations, we train a transformer \cite{attentionisallyouneed, t5} model on the task of multi-track MIDI infilling. 
%resulting in a model that can help composers flesh out, continue, and/or create variations within their compositions. Our approach 
Our model allows composers to generate new notes for arbitrary subsets of track-measures in their compositions, conditioned on any contiguous slice of measures containing the subset. (By a {\em track-measure}, we simply mean a measure within a track---see Figure \ref{FigPrompt}.) 
%Motivated primarily by our experience trying to use these models as compositional aides, we apply transformers \cite{attentionisallyouneed, t5} to the task of multi-track MIDI infilling, resulting in a trained model that can help composers flesh out, continue, and/or create variations within their compositions. Our approach allows composers to generate new notes for arbitrary subsets of track-measures in their compositions, conditioned on any contiguous slice of measures containing the subset. (A {\em track-measure} is a measure within a track---see Figure \ref{FigPrompt}.) 
We also build a novel system for interacting with our model in the REAPER digital audio workstation (DAW).\footnoteone Our system is cross-platform and easy to install.
%(requiring the user only to have \tt{python} \cite{python}, \tt{pytorch} \cite{pytorch}, and the Hugging Face \tt{transformers} library \cite{huggingface} installed), 
%and enables rapid interaction with the model. 
When a user runs one of our REAPER scripts, a model prompt is built directly from the slice of measures selected in the user's REAPER project, our model evaluates the prompt, and the model output is written back into the user's project---see Figure \ref{FigPrompt}. All of this happens within a few seconds, allowing the user to listen to the output, modify it to create a new prompt, generate an output from that, etc. This allows our model to be used in an interactive manner, where model outputs are refined by the user over the course of several prompts.

We note that our infilling objective includes continuing a piece of music, simply by including empty measures at the end of the piece in the prompt. Additionally, our model has the ability to write variations: One can randomly mask $1/n$ of the track-measures in a measure selection and ask the model to fill in those parts, then feed the result back into the model with another $1/n$ masked, and so on, until all track-measures have been masked and filled.

Toward the end of this project we discovered MMM \cite{MMM, MMM2}, which consists of two separate GPT-2--like \cite{gpt2} models trained on the tasks of measure infilling and track infilling. The authors include code to use these models to infill arbitrary subsets of track-measures, as we do. MMM comes in 4-bar and 8-bar variants, which are limited to inputs with 12 and 6 tracks, respectively, and its web demo is limited to inputs having a 4/4 time signature.
%, and the authors do not evaluate their models. 
%Since MMM is the closest to our work, we compare our model to the MMM models in Sections \ref{SecTestProcedure}--\ref{SecSubjective}.

The primary contributions of this work are as follows. First, in Section \ref{SecData} we introduce a novel data filtering and preprocessing approach, applicable to any MIDI dataset used for training models. Our approach helps rectify certain issues we have encountered when using other models. Second, we train and release a new model, capable of infilling arbitrary track-measures in an arbitrary slice of measures in a MIDI file, with no effective restriction (aside from a soft input token limit of 1650) on tempos, number of measures, or number of instrument tracks. Tracks may be polyphonic or monophonic in any combination. The only time signature restriction is that all measures must be eight quarter notes or fewer. 
%We use a temporal resolution that accommodates 32nd notes and 16th note triplets. 
%, which we have found sufficient for expressive generation. 
Our model is more flexible than MMM and compares favorably to MMM in both objective and subjective tests---see Sections \ref{SecObjective}--\ref{SecSubjective}. Additionally, our model was trained only on permissively-licensed MIDI files, so its outputs should be usable by composers with minimal risk---see Section \ref{SecModel}. Finally, we release our complete system, including training code and scripts that enable rapid interaction with our model in REAPER. Our model is the first DAW-integrated model capable of infilling parts for all 128 pitched MIDI instruments (including repeated instruments) and drums, in any combination.

\section{Related Work} 
\label{SecRelatedWork}

As mentioned in Section \ref{SecIntro}, MMM \cite{MMM, MMM2} performs multi-track infilling for all MIDI instruments (subject to bar and track limits), and DeepBach \cite{deepbach} performs multi-track infilling for Bach-like chorales. Coconet \cite{coconet} also performs multi-track infilling for Bach-like chorales. MusIAC \cite{musiac} incorporates user controls and performs track-based and measure-based infilling, although its inputs and outputs are limited to a maximum of three tracks,
%(one melody track, one bass track, and one accompaniment track, with no drums), 
16 measures, and four common time signatures.
%, and a temporal resolution of 16th notes. 
MusicVAE \cite{musicvae} can interpolate between two given clips of music, which can be viewed as a type of infilling.
To our knowledge, all other existing music infilling systems are limited to monophonic infilling \cite{diffusion, phraseinpainting, traverselatent, sketchnet, drawandlisten} or single-instrument infilling \cite{pianoinpainting, infill-xlnet}. 

Generating or continuing a piece of music can be seen as a special case of infilling. 
%To generate a piece from scratch we  simply prompt our model with a set of empty measures, and to continue a piece of music we prompt our model with a piece of music with a set of empty measures at the end of it. 
Models which can generate or continue a piece of music include \cite{musenet, symphonynet, musictransformer, pmt, cwt}.
%Pop Music Transformer \cite{pmt} and Compound Word Transformer \cite{cwt}.

Previous DAW-integrated generative music systems include \cite{deepbach, pianoinpainting, magentastudio}. 
%Previous models with DAW integration include DeepBach \cite{deepbach}, the Piano Inpainting Application \cite{pianoinpainting}, and Magenta Studio \cite{magentastudio}. 
NONOTO \cite{nonoto} is a model-agnostic interface that can be linked with a model to perform interactive measure-based infilling. This interface could potentially be altered to allow for the expanded type of infilling our model is capable of. However, we opt to build an interface between our model and REAPER directly, essentially using REAPER as the GUI for our model.
%NONOTO \cite{nonoto} is a web-based interface that can be used to link a model with a DAW to perform interactive measure-based infilling. This interface could potentially be altered to allow for the expanded type of infilling our model is capable of. However, we opt to build an interface between our model and REAPER directly, essentially using REAPER as the GUI for our model.

\section{Data Filtering and Preprocessing}
\label{SecData}
In this section we describe our filtering and preprocessing approach, any portion of which can be applied to any dataset of MIDI files.
First, we remove from the dataset any file whose notes seem to have no relation to the underlying grid (Section \ref{SecCosSim}). Next, we dedupe files from the dataset using note onset chromagrams (Section \ref{SecDedupe}). Finally, we preprocess all remaining files to standardize properties like track order (Section \ref{SecPreprocessing}). This final preprocessing step includes a method for detecting and removing shifted duplicate and near-duplicate tracks within files (Section \ref{SecShiftDetection}). 

\subsection{Cosine Similarity for On-Grid Note Detection} 
\label{SecCosSim}

Every MIDI file has a measure and grid structure defined by tempo and time signature events. 
%The default tempo is 120 beats per minute (BPM) and the default time signature is 4/4. 
However,  MIDI file authors are free to ignore this structure, and frequently do when recording free-flowing performances.
Other models we have used occasionally write a note in the wrong place---e.g., a 32nd note away from where it clearly should be---and a small experiment we ran suggests that training on MIDI files that don't quantize well to the grid used by the model is a major cause of this. To address this, 
%Since measure and position-within-measure information is important for our modeling, 
we remove from our dataset any MIDI file whose note onsets appear to have no relation to the underlying grid. This is not as simple as checking whether all (or most) note onsets occur on the grid, as many MIDI file authors who use the grid include ``humanization'' of note timings, where many note onsets that occur slightly off the grid nevertheless quantize correctly to the grid. For instance, in a MIDI file with a resolution of 960 ticks per quarter note, a humanized quarter-note performance might have notes occurring in a 40-tick window centered at every $960$th tick.

To perform this filtering, given a MIDI file $M$, we quantize the note onsets in $M$ to a resolution of 12 ticks per quarter note, and we form a length-12 vector $v_M$ whose $i$th entry ($i\in\{0 ,\ldots, 11\}$) is the number of note onsets in $M$ occurring $i$ ticks after a grid quarter note. The idea is that if the note onsets in $M$ have nothing to do with the grid, then $v_M$ will point in a similar direction to the uniform vector $v_1 = (1, \ldots, 1)\in \R^{12}$. We therefore compute the cosine of the angle $\theta_M$ between $v_M$ and $v_1$:
\[
\cos(\theta_M) = \frac{\langle v_M, v_1 \rangle}{||v_M|| \cdot ||v_1||},
\]
and we declare a threshold $T$ such that when $\cos(\theta_M) > T$ we remove the file $M$ from our dataset. 
Hand exploration indicated that $T = 0.8$ was a reasonable threshold, which we chose for this project. 
We note that a straight fully-on-grid 8th-note pattern $M$ has $\cos(\theta_M) \approx 0.41$ and a straight fully-on-grid 16th-note pattern $M$ has $\cos(\theta_M) \approx 0.58$. 
%We also note that the crude resolution of 12 ticks per quarter note used in this section does not exclude files that quantize well to a finer-resolution grid. The approach taken in this section merely removes files that generally don't respect downbeats.

\subsection{Deduping Using Note Onset Chromagrams} 
\label{SecDedupe}

We dedupe our dataset to avoid data imbalance during training and to prevent overlap between our training and test sets. Given a MIDI file $M$, we compute a size-12 set of note onset chromagrams using the following procedure. 

First, we remove all drum tracks from $M$.
Then, using a 12-tick-per-quarter-note grid, we quantize the note onsets in $M$ to the nearest 16th note or 8th note triplet.
 %on a grid that contains 12 ticks per quarter note.
%which contains 12 ticks per quarter note.
%, which consists of note onset locations for 16th notes and 8th note triplets. This grid contains 12 evenly-spaced ticks per quarter note, of which 6 are valid locations for note onsets. 
Then, we remove all empty measures at the beginning and end of $M$, and we replace each set of contiguous empty measures within $M$ with one empty measure. Then, for each tick in $M$ and for each pitch class,
%(a pitch class is just an equivalence class of integers mod 12),
we record whether $M$ has at least one note onset of that pitch class at that tick. This information comprises one note onset chromagram for $M$. The other 11 come from repeating this procedure for each possible transposition of $M$. We dedupe the dataset by keeping only one file with a given set of note onset chromagrams. Quantization helps us catch files that differ only trivially in grid resolution and/or note onset times, while transposition helps us catch files that differ only in key.

\subsection{Preprocessing of Individual MIDI Files}
\label{SecPreprocessing}

After deduping, we preprocess each MIDI file in our dataset in the following way. 

First, we arrange the information in the MIDI file so that each track holds notes for one instrument. We order tracks according to their MIDI instrument number (0--127), taking drums as instrument 128. We also consolidate all drum tracks to a single track, and we apply a drum simplification map (consolidating, e.g., three different bass drum pitches to a single pitch).

Next, we apply pedal information in the file (if present) to extend note lengths, and then delete all continuous controller (cc) data. We do not model cc data in this project.

With the exception of drums, we allow multiple tracks to use the same instrument. 
However, when this happens, if there is more than one track having a given instrument, we remove all but one of those tracks that are equal to, a shift of, or close to a shift of another track with the same instrument, using the procedure in Section \ref{SecShiftDetection}. 

We impose the restriction that all measures must be eight quarter notes or fewer. If any time signature in the file declares longer measures, we alter the time signatures to shorten the measures. 

Finally, using a 24-tick-per-quarter-note grid, we quantize the events in the file to the nearest 32nd note or 16th note triplet. This is ultimately the level of quantization we use to train our model. (Earlier experiments involved quantizing to 16th notes or 16th notes + 8th note triplets, which we found insufficient for expressive generation.)

\subsection{Removing Shifted Duplicate and Near-Duplicate Tracks}
\label{SecShiftDetection}
A MIDI file may contain duplicate tracks. Such tracks contain no useful information for modeling, so we remove them. Shifted duplicate tracks are frequently used by MIDI file authors to encode {\em delay} effects (as the MIDI spec offers no way to encode the use of a delay directly). Choosing to use a delay is a mixing decision, not a compositional decision, and we want our model to focus on making compositional decisions, so we remove shifted duplicate tracks as well. We have also seen tracks that are duplicates or shifted duplicates of other tracks within a file, plus or minus a few notes and/or humanization. We remove such near-duplicate tracks as well.

Given a note $n$ in a track $T$, let $\start(n)$ and $\nend(n)$ indicate the start and end times of the note $n$, respectively, and let $\pitch(n)\in\{0 ,\ldots, 127\}$ indicate the MIDI pitch of $n$. We record, for $p\in \{0 ,\ldots, 127\}$, the union of closed intervals
\[
I_T(p) = \cup_{\up{n} \in T: \pitch(n)=p} \{[\start(n), \nend(n)]\} \subseteq \R,
\]
and we define
$
|I_T| = \sum_{p=0}^{127} |I_T(p)|,
$
where $|I_T(p)|$ is the sum of the lengths of the disjoint intervals in $I_T(p)$.

Given tracks $T_1$ and $T_2$, we define the {\em overlap measure} $O(T_1, T_2)\in [0, 1] \subseteq \R$ to be
\[
O(T_1, T_2) = \frac{\sum_{p=0}^{127}|I_{T_1}(p) \cap I_{T_2}(p)|}{\max\left(|I_{T_1}|, |I_{T_2}| \right)} .
\]
The idea is that $O(T_1, T_2)$ measures the percentage of the note intervals in the larger of the two tracks accounted for by the note intervals in the smaller of the two.

We use a threshold of 0.9 for asserting near-overlap between two tracks. As we go through the tracks in a MIDI file in order, a later track $T$ is thrown out if there exists an earlier track $T_0$ using the same instrument such that some shift $T_s$ of $T$ of no more than a half note has the property that $O(T_0, T_s) \geq 0.9$. In our experience with our trained model, we have found this preprocessing step sufficient to prevent the model from outputting duplicates or shifted duplicates of tracks in its inputs.

%Second, generative language models often fall into states where they simply copy information from their prompts to their outputs. We want to discourage this as much as possible by not including examples in our training dataset where that is the ``right" thing to do.
%Third, users of our models would be disappointed if the models frequently output copies of information they were prompted with. The point of our models is to generate new and different musical ideas that accompany the ideas they were prompted with.

\section{Our Language}
\label{SecLanguage}
\label{SecLanguage1}

After applying the procedure from Section \ref{SecData} to a collection of MIDI files, we process the files into an event-based language for modeling.
%In fact, we use two languages: A basic language consisting of MIDI-like event instructions (Section \ref{SecLanguage1}) and a SentencePiece \cite{spm} unigram language learned from our training data that joins common contiguous event instructions into single tokens (Section \ref{SecLanguage2}). 
Our language is similar to the standard event-based MIDI language used for piano performance modeling in \cite{musictransformer}. However, we use explicit measure tokens to denote the start of each measure. Also, we do not model velocity of individual notes directly. Instead, we assign a dynamics level to each measure based on the average velocity of the notes in the measure. We use eight dynamics levels, with thresholds learned from data.

%%%%%%%%%%%%%%%%%%%%%%%%%%%

%%%%%%%%%%%%%%%%%%%%%%%%%%%%%%%%%%%

The tokens used by our language are as follows:
\begin{itemize}
	\item  M:$x$, $x\in\{0 ,\ldots, 7\}$. Declares a measure of dynamics level $x$. 
	\item B:$x$, $x\in\{0 ,\ldots, 7\}$. Declares the tempo (BPM) level at the start of a measure. We use eight tempo levels, with thresholds learned from data. 
	\item L:$x$, $x\in\{1 ,\ldots, 192\}$. Declares that a measure has length equal to $x$ ticks.
	\item I:$x$, $x\in\{0 ,\ldots, 128\}$. Changes the current instrument to MIDI instrument $x$ ($128 = \up{drums}$).
	\item R:$x$, $x\in\{1 ,\ldots, 63\}$. Declares that the current instrument is the same MIDI instrument as another instrument in the file/project, but on a different track. Higher $x$ values indicate lower average pitch.
	\item N:$x$, $x\in\{0 ,\ldots, 127\}$. Note of pitch $x$. Used by instruments 0--127.
	\item D:$x$, $x\in\{0 ,\ldots, 127\}$. Drum hit of drum pitch $x$. Used by instrument 128.
	\item d:$x$, $x\in\{0 ,\ldots, 192\}$. Sets the duration of each note declared from this point forward to $x$ ticks.
	\item w:$x$, $x\in\{1 ,\ldots, 191\}$. Advances the current insertion point for new notes in the measure by $x$ ticks.
	\item \tt{<extra\_id\_x>}, $x\in\{0 ,\ldots, 255\}$. Mask tokens.
	\item \tt{<mono>}, \tt{<poly>}. Instructs the model to write a monophonic or polyphonic part for a masked part. For our purposes a monophonic part is one where no two notes in the part have the same onset tick.
\end{itemize}

%%%%%%%%%%%%%%%%%%%%%%%%%%%%%%%%%%%

\begin{figure}[t]
 \centerline{\framebox{
 \includegraphics[width=0.9\columnwidth]{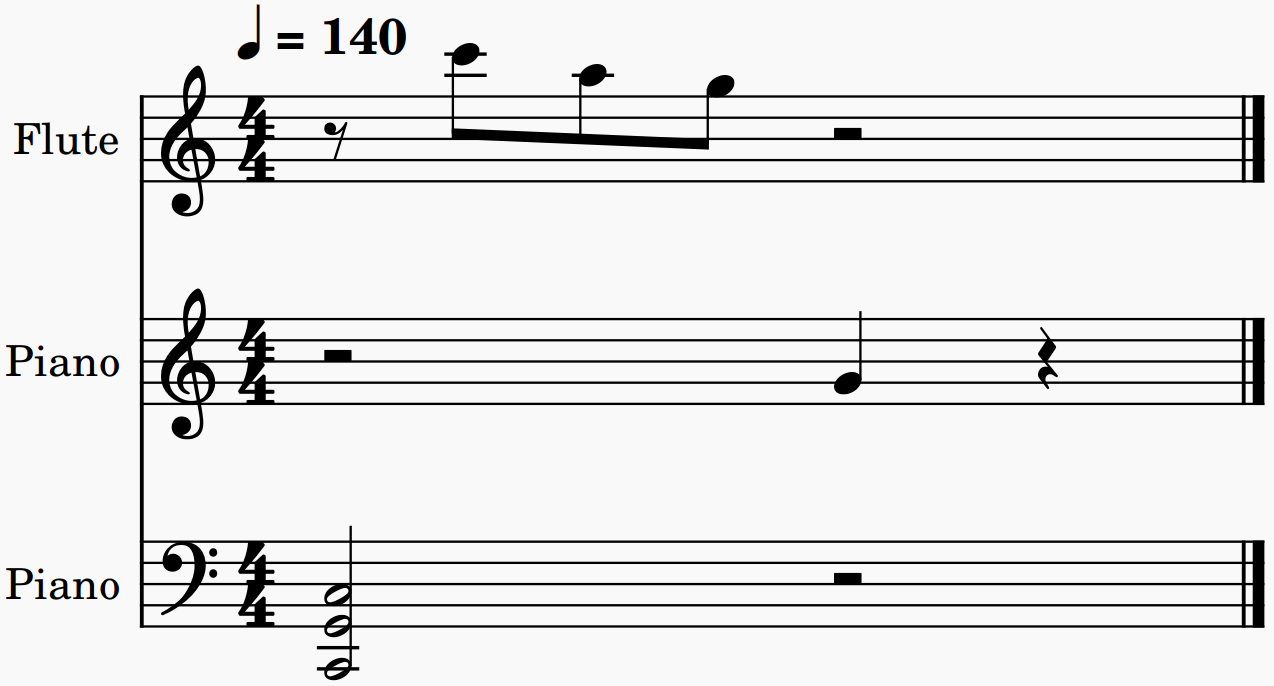}}}
 \caption{We tokenize this measure as M:5 B:6 L:96 I:0 w:48 d:24 N:67 I:0 R:1 d:48 N:36 N:43 N:48 I:73 w:12 d:12 N:84 w:12 N:81 w:12 N:79. 
Note that piano and flute are MIDI instruments 0 and 73.
}
 \label{FigVocabEx}
\end{figure}

Tokens are assembled in a standardized manner to represent measures. 
Each measure begins with M:$x$, B:$x$, and L:$x$ tokens. 
I: commands are included for a measure only when that instrument is present in the measure. We do not intermingle instrument note instructions as we write each measure from left to right (as MuseNet \cite{musenet} did), as that would make it difficult to mask individual instrument parts within measures. Rather, we write the full part for one instrument within the measure before writing the full part for the next instrument within the measure. Figure \ref{FigVocabEx} contains an example of a tokenized measure. 

To form songs, we simply concatenate measures.

\section{Model, Data, and Training Procedure}
\label{SecModel}
\label{SecTraining}

We use recent recommendations from the language modeling community to design and train our model. Based on the recommendations in \cite{pegasusx, t511, scalinglaws}, we choose a T5 (full, relative-attention) encoder-decoder architecture  \cite{t5}.
%with gated GELU activation function. 
We opt for a full attention model because such models were found to outperform memory-efficient models in \cite{pegasusx} when the full input sequence fits in memory, as we expect to be the case in most real-world applications of our model. Also, we adopt the {\em DeepNarrow} strategy of \cite{scaleefficiently}, opting for a model dimension of 384, 10 encoder layers, and 10 decoder layers. 
%These parameters were carefully chosen for reasons described in Section \ref{SecRisks}. 
For training, we use the \tt{pytorch} \cite{pytorch} Hugging Face \cite{huggingface} implementation of T5.
For inference, we use nucleus sampling \cite{nucleussampling} with a threshold of $p=0.95$. 

To train a model that is essentially free of copyright worry, we collect MIDI files from the internet marked as being in the public domain, freely available to use without attribution, or available under a CC BY license. 
We exclude files marked as having share-alike or non-commercial licenses, since we want composers to be able to use model outputs however they wish. 
We also collect private donations and files from the internet for which we secure direct permission from the MIDI file authors to use for training. This results in a dataset of approximately 40K MIDI files after filtering. Most of our training files are in Western classical, folk, and hymnal styles, although some modern styles are also present. 
%Sources for most of the files we used for training are available in the acknowledgments in our GitHub repository.\footnotetwo 

%\subsection{Pretraining} 
\label{SecPT}
We follow the standard approach to the training of large language models of splitting our training procedure into pretraining and finetuning phases. A similar approach was also used in \cite{lakhnes}.
For pretraining, we use the T5 corrupted-span sequence-to-sequence objective \cite{t5}. We begin by pretraining on the 192K training files in the CocoChorales \cite{coco} dataset and their piano reductions for three epochs. The CocoChorales are only used for this initial pretraining to teach the model the basics of music theory and our language. We then move on to our dataset of 40K MIDI files. After tokenization and corruption, we greedily chunk each song into inputs of 512 or fewer (short) or 1650 or fewer (long) tokens. 
Additionally, each song in our dataset is transposed a random amount between -5 and +6 semitones (inclusive) for each epoch. 
Following the recommendations in \cite{pegasusx}, we train our model on short examples for 20 epochs and then long examples for 11 epochs. We release the resulting pretrained model, which others may find useful for finetuning on downstream tasks. 
%As in \cite{t511}, we use a dropout rate of 0 for pretraining.

%\subsection{Finetuning} 
\label{SecFT}
For finetuning, we continue to leverage the corrupted-span sequence-to-sequence objective to finetune our model on the task of multi-track MIDI infilling. We create training examples from songs in our training dataset by taking slices of measures from the songs and masking subsets of track-measures from these slices. 
During finetuning every N:, D:, d:, and w: token for a given track-measure is masked, and corresponds to a single mask token. 
%In contrast to pretraining, where any tokens can be masked and each mask token corresponds to only a few unmasked tokens, during finetuning every N:, D:, d:, and w: token for a given track-measure is masked and corresponds to a single mask token. 
With probability 0.75, we add a \tt{<poly>} or \tt{<mono>} token corresponding to the nature of the masked tokens for each mask. (We choose not to include these tokens in every training example since users will not always include these instructions in their prompts.)
%We train on an NVIDIA 12 GB P100 accelerator, and 
%limit our examples to 1650 tokens or less for each input and output. 
Finetuning examples are limited to inputs with a maximum of 1650 tokens and outputs with a maximum of 1650 tokens.

We generate our finetuning masks by selecting from mask patterns that we consider to be musically relevant and/or useful for training. 
%(See Table ??? in Appendix \ref{AppendixMaskPatterns}.) 
To help train our model for use on small numbers of measures, we also occasionally (15\% of the time) truncate examples to a random smaller number of measures than the number allowed by our token limits. As with pretraining, each example is transposed randomly. 
We finetune our model for 51 epochs, and we release the resulting finetuned model.
%We use a dropout rate of 0.1 (10\%) during finetuning. 

\begin{table*}[!ht]
\begin{center}
\begin{tabular}{|l|c|c|c|c|}
\hline
Task & Our Model & Our Model -MP & MMM-8 & MMM-4 \\
\hline
\multicolumn{5}{c}{Note $F_1$ results. Higher is better.}\\
\hline
8-bar random infill & {\bf 0.5414} $\pm$ $(0.1887)^a$ & 0.5315 $\pm$ $(0.1904)^b$ & 0.4153 $\pm$ $(0.1819)^c$ & 0.4025 $\pm$ $(0.1652)^d$\\
\hline
16-bar random infill $^\ast$ & {\bf 0.5771} $\pm$ $(0.1661)^a$ & 0.5705 $\pm$ $(0.1669)^b$ & 0.4133 $\pm$ $(0.1534)^c$ & 0.4059 $\pm$ $(0.1399)^d$\\
\hline
8-bar track infill & {\bf 0.179} $\pm$ $(0.1902)^a$ & 0.1634 $\pm$ $(0.18)^b$ & 0.1063 $\pm$ $(0.1573)^d$ & 0.1427 $\pm$ $(0.1646)^c$\\
\hline
16-bar track infill & {\bf 0.1773} $\pm$ $(0.1752)^a$ & 0.1609 $\pm$ $(0.165)^b$ & 0.1107 $\pm$ $(0.1383)^d$ & 0.1467 $\pm$ $(0.1489)^c$\\
\hline
8-bar last-bar fill & 0.5019 $\pm$ $(0.2719)^a$ & {\bf 0.5063} $\pm$ $(0.2751)^a$ & 0.4329 $\pm$ $(0.2445)^b$ & 0.3756 $\pm$ $(0.2289)^c$\\
\hline
16-bar last-bar fill $^\ast$ & {\bf 0.5415} $\pm$ $(0.2853)^a$ & 0.539 $\pm$ $(0.2823)^a$ & 0.4338 $\pm$ $(0.2468)^b$ & 0.3818 $\pm$ $(0.2293)^c$\\
\hline
\multicolumn{5}{c}{Pitch class histogram entropy difference results. Lower is better.}\\
\hline
8-bar random infill & {\bf 0.2845} $\pm$ $(0.1627)^a$ & 0.2948 $\pm$ $(0.1597)^b$ & 0.3045 $\pm$ $(0.1561)^c$ & 0.3049 $\pm$ $(0.1497)^c$\\
\hline
16-bar random infill & {\bf 0.2691} $\pm$ $(0.1325)^a$ & 0.2797 $\pm$ $(0.1326)^b$ & 0.3093 $\pm$ $(0.124)^c$ & 0.3063 $\pm$ $(0.1138)^c$\\
\hline
8-bar track infill & 0.3933 $\pm$ $(0.3032)^c$ & 0.42 $\pm$ $(0.3134)^d$ & {\bf 0.2864} $\pm$ $(0.2966)^a$ & 0.3021 $\pm$ $(0.2517)^b$\\
\hline
16-bar track infill & 0.3842 $\pm$ $(0.2654)^c$ & 0.3995 $\pm$ $(0.2763)^c$ & {\bf 0.284} $\pm$ $(0.2348)^a$ & 0.3036 $\pm$ $(0.2072)^b$\\
\hline
8-bar last-bar fill & {\bf 0.3018} $\pm$ $(0.2661)^a$ & 0.3072 $\pm$ $(0.2692)^a$ & 0.3213 $\pm$ $(0.2602)^b$ & 0.3439 $\pm$ $(0.2777)^c$\\
\hline
16-bar last-bar fill $^\ast$ & {\bf 0.2851} $\pm$ $(0.2652)^a$ & 0.2925 $\pm$ $(0.2672)^a$ & 0.3209 $\pm$ $(0.2619)^b$ & 0.3454 $\pm$ $(0.2741)^c$\\
\hline
\multicolumn{5}{c}{Groove similarity results. Higher is better.}\\
\hline
8-bar random infill & {\bf 0.9534} $\pm$ $(0.0298)^a$ & 0.9519 $\pm$ $(0.0306)^b$ & 0.9333 $\pm$ $(0.0369)^c$ & 0.9314 $\pm$ $(0.0364)^d$\\
\hline
16-bar random infill $^\ast$ & {\bf 0.956} $\pm$ $(0.027)^a$ & 0.9552 $\pm$ $(0.0275)^b$ & 0.9323 $\pm$ $(0.0337)^c$ & 0.9317 $\pm$ $(0.032)^c$\\
\hline
8-bar track infill & {\bf 0.9115} $\pm$ $(0.0592)^a$ & 0.9069 $\pm$ $(0.0617)^b$ & 0.8921 $\pm$ $(0.0695)^d$ & 0.8987 $\pm$ $(0.0626)^c$\\
\hline
16-bar track infill & {\bf 0.9113} $\pm$ $(0.0547)^a$ & 0.9082 $\pm$ $(0.0553)^b$ & 0.8946 $\pm$ $(0.0561)^d$ & 0.9011 $\pm$ $(0.0536)^c$\\
\hline
8-bar last-bar fill & 0.9517 $\pm$ $(0.0414)^a$ & {\bf 0.9524} $\pm$ $(0.0411)^a$ & 0.9381 $\pm$ $(0.045)^b$ & 0.9334 $\pm$ $(0.0457)^c$\\
\hline
16-bar last-bar fill $^\ast$ & {\bf 0.9544} $\pm$ $(0.0481)^a$ & 0.9542 $\pm$ $(0.0424)^a$ & 0.938 $\pm$ $(0.051)^b$ & 0.9339 $\pm$ $(0.0475)^c$\\
\hline
\end{tabular}
\end{center}
\caption{Objective infilling summary statistics. All cells are of the form mean $\pm$ (std dev)$^s$, where $s$ is a letter. Different letters within a row indicate significant location differences ($p < 0.01$) in the samples for that row
according to a Wilcoxon signed rank test with Holm-Bonferroni correction. 
Asterisks ($\ast$) indicate a significant performance difference ($p < 0.01$) between a 16-bar task and the 8-bar task in the previous row for our model according to a Wilcoxon rank sum test.} 
\label{TableObjectiveResults}
\end{table*}

\section{Objective Evaluation of our Model}
\label{SecTestProcedure}
\label{SecObjective}

To form our test set, we select a set of 2500 MIDI files from the Lakh MIDI dataset \cite{lmd2, lmd1} that is disjoint (according to the procedure in Section \ref{SecDedupe}) from our training set, all 
in 4/4 time and 
having at least 16 measures.
%, and passing our test in Section \ref{SecCosSim} for on-grid note detection. 
Given a MIDI file in our test set, for each of the three mask patterns below, we select an 8- and a 16-measure slice of the file and mask the selected slice with that mask pattern. We thus generate six test examples from each test file, corresponding to the six different tasks on which we evaluate models. 
Given a slice of measures, our mask patterns for testing are:
\begin{enumerate}
	\item Random mask: Each track-measure in the slice is masked with probability 0.5.
	\item Track mask: Up to half of the tracks $t$ are selected at random from the slice, and every measure of each such track $t$ is masked.
	\item Last-bar mask: Given the last measure $m$ of the slice, measure $m$ of every track is masked. This pattern is used to measure the ability of models to continue songs.
\end{enumerate}
%Of course, we ensure that every test example has at least some note information in both the input and the expected output.
The ground truth for each example consists of the masked notes in the example.
In our test data, 99\% and 75\% of our 8-measure and 16-measure prompts (respectively) encode to 1650 or fewer tokens.
When input prompts are longer than 1650 tokens, we chunk the prompts prior to evaluating them with our model. 
%, and then reassemble the model outputs. 
%For our measure-masked prompts, we select as many measures as possible for the last chunk and keep only that (as it is the only relevant chunk for those prompts). For other patterns, we greedily select as many measures as possible for each chunk except for the last two chunks, which are kept to as close of a size 
% (in terms of # of tokens)
%as possible. 
%We do this because we want to avoid evaluating short ending chunks that might not contain enough information for infilling, which could unfairly negatively affect our results for examples that require chunking. 
%To prevent out-of-memory errors, outputs (per chunk) are limited to a maximum of 3000 tokens. None of our chunks have intended outputs with more than 3000 tokens.  - sentence not needed anymore

To compare our model to MMM \cite{MMM, MMM2}, we modify the MMM Colab worksheet to run our examples through the MMM models in batches. We recreate our test examples, quantizing them from their underlying MIDI files to MMM's 12-tick-per-quarter-note resolution, and then modify them to accommodate the restrictions of the MMM models: Since the 4-bar and 8-bar MMM models are limited to inputs containing a maximum of 12 and 6 tracks, respectively, we chunk each test example into 4-bar and 8-bar chunks, and then we split each chunk into sub-chunks consisting of up to 12 and 6 tracks.
%, and reassembled the outputs afterward. 
The MMM models have a strict input + output token limit of 2048, so when sub-chunking, we only add enough tracks to a sub-chunk to ensure that the input + ground truth has no more than 2048 tokens. This biases the comparison in favor of the MMM models somewhat, as this requires us to look at the length of the ground truth as part of the input chunking procedure. 
%In contrast, we did not use the intended outputs in any way to chunk our examples for our models. 
Also, our test set is contained in MMM's training set, but there is no reasonable way to avoid this as the MMM models were trained on the full Lakh MIDI dataset. (We wanted a diverse and well-randomized test set, and the Lakh MIDI dataset is the only publicly-available dataset we are aware of that fits this bill.)

We evaluate models with standard metrics: Note $F_1$ \cite{mt3}, average pitch class histogram entropy difference \cite{infill-xlnet, jazztransformer}, and average groove similarity \cite{infill-xlnet, jazztransformer}. Note $F_1$ measures how closely the generated notes match, on a note-for-note basis, the notes in the ground truth. (For our purposes, a generated note matches a note $n$ in the ground truth if and only if its onset tick, measure, pitch, and track match exactly those of $n$.) The other metrics measure how well certain higher-level statistics of the generated notes match those of real music. 
For pitch class histogram entropy calculations, drums are ignored. 
Each metric is computed on a per-example basis, and then for each model, task, and metric, the 2500 results are averaged to give the results in Table \ref{TableObjectiveResults}.
%, and is then averaged across the examples for that task (as we consider each example to be equally important). 
For fairness of groove similarity comparison, we use a denominator of 48 for all models. (This is reasonable, as our models and the MMM models both effectively have 48 possible note onset positions per 4/4 measure.) For our model, ``-MP'' indicates that the examples were evaluated without \tt{<mono>} or \tt{<poly>} tokens present.

%To explain how we calculate note $F_1$ for a test example, each generated note $n$ is for a particular track index $n$.tr (recall that an instrument may appear on multiple tracks) and a particular measure $n$.measure, and has pitch notated by $n$.pitch and onset time within its measure notated by $n$.tick. Viewing infilling as a document retrieval task, we consider the set of generated tuples 
%\[
%S_{ret} = \{(n.\up{tr}, n.\up{measure}, n.\up{pitch}, n.\up{tick}): n \up{ generated}\}
%\]
%to be the set of {\em retrieved documents}, and we consider the corresponding set of 4-tuples from the ground truth to be the set $S_{rel}$ of {\em relevant documents}. Tuples must match exactly to be considered a match. The {\em precision} of the output is $|S_{ret} \cap S_{rel}| / |S_{ret}|$, while the {\em recall} of the output is $|S_{ret} \cap S_{rel}| / |S_{rel}|$.
%$F_1$ is computed in the standard way from precision and recall: 
%\[
%F_1 = 2 \frac{\up{precision} \cdot \up{recall}}{\up{precision} + \up{recall}}.
%\]
%The note $F_1$ score for any example lies between 0.0 (if either precision or recall is 0) and 1.0 (if precision and recall are both perfect).

For each row of Table \ref{TableObjectiveResults}, we perform a Wilcoxon signed rank test \cite{wilcoxon} with Holm-Bonferroni correction \cite{holm}. We find significant differences between our model and the MMM models in all 18 rows, with our model outperforming the MMM models in 16 out of 18 rows. The MMM models outperform our model only for pitch class histogram entropy difference for full-track infilling. 

Additionally, we find a significant difference in our model's performance when \tt{<mono>} and \tt{<poly>} tokens are included in prompts in 11 out of 18 rows. All significant differences favor including these tokens, suggesting that the development of additional user controls (as in \cite{musiac}) would be a useful line of future work.
%, and provides a baseline for comparison. 

Finally, a Wilcoxon rank sum test \cite{wilcoxon} reveals significant differences ($p<0.01$) in 8-bar versus 16-bar results for our model in five out of nine comparisons. All significant differences favor the 16-bar results, emphasizing the importance of training on longer measure slices. However, we never observe a significant difference in 8-bar versus 16-bar results for track infilling, suggesting that larger context windows generally provide no additional useful information for completing this particular task. 

Additional experiments not reported here indicate that scaling our training approach (training larger models on more data) is a feasible path for improving model performance on the metrics presented here.

%In another experiment, we trained a larger Composer's Assistant model on a larger dataset (specifically, the union of the Lakh MIDI dataset \cite{lmd1, lmd2}, the SymphonyNet dataset \cite{symphonynet}, and the VGMusic dataset \cite{VGMusic}) and obtained significantly better metrics, indicating that scaling our training approach is feasible.

%Also, while precision and recall are not reported directly in the table, precision and recall averages are similar to the $F_1$ averages reported in this table. 

\section{Subjective Evaluation of our Model}
\label{SecSubjective}

\begin{table}
 \begin{center}
 \begin{tabular}{|l|c|c|c|}
  \hline
  & Real Music & Our Model & MMM \\
  \hline
	1st place & \textbf{66} & 32 & 27\\
	\hline
	Avg rank & \textbf{1.664} & 2.032 & 2.304\\
	\hline
 \end{tabular}

 \begin{tabular}{|l|c|c|}
  \hline
  $p$-values & Our Model & MMM \\
  \hline
	MMM & 0.0239 & - \\
	\hline
	Real Music & 0.0034 & $2.3\cdot 10^{-5}$\\
	\hline
 \end{tabular}

\end{center}
 \caption{Subjective results from our listening test.}
 \label{TableSubjectiveTest}
\end{table}

While the results in Section \ref{SecTestProcedure} are encouraging, the ground truth may not reflect the only reasonable way to fill in missing notes. To help address this, we conducted a small listening test with 25 participants. 
%Our model and MMM both generally create rhythmically fluent outputs, so we did not feel the need to test for rhythmic fluency. 
We prepared nine examples mostly involving melodic generation.
%, which is the task we have observed our models to be weakest at and also the task that is easiest for listeners to hear differences on. 
Each example consisted of three 8-measure clips, one of which was real multi-track music. The other two clips were created by removing some tracks from the real music and using our model and MMM to fill those tracks. Participants were shown five of the nine examples at random, and for each example were asked to rank the three clips in order of preference. Results are given in Table \ref{TableSubjectiveTest}. 

A Wilcoxon signed rank test with Holm-Bonferroni correction reveals significant differences in rankings between all three types of music, with $p$-values given in Table \ref{TableSubjectiveTest}. In this test we see a clear preference for real music, and a significant ($p<0.05$) preference for music generated by our model over music generated by MMM. 
One expert participant 
%(who ranked real music in first place in four out of five examples) 
commented that melodies generated by the models were generally more directionless than those in real music, often failing to drive towards a cadence or ``payoff.'' 
%This coincides with our experience using our model: When using it as a compositional aid, we often have to alter (usually by simplifying) the ends of melodic phrases our model generates. 
We agree with this assessment, and 
this is a shortcoming of our model that we hope to address in future work.

\section{Limitations and Risks}
\label{SecLimitations}
\label{SecRisks}
Our model writes music that is reflective of its training set. Most of our training files are in Western classical, folk, and hymnal styles.
%We note that all standard MIDI instruments, including drums, are present in our training set, but at imbalanced proportions. 
%Since we do not model pitch bends, our model is not capable of understanding or writing microtonal music. 
While we included in our training set only files marked as being permissively licensed, it is possible that some files were mismarked. It is also theoretically possible for our model to output copyrighted music, even if such music was not present in the training set.

The most common request we have heard from composers to whom we have shown our system is {\em personalization}. Generally speaking, they do not want systems that write full songs, and they do not want systems that write ``generic'' music. Rather, they want systems that can suggest ideas in their style. Some small experiments indicate that our finetuned model can be personalized by individuals (by continuing to finetune the model on their own MIDI files) to write in their styles. Low-rank adaptation \cite{lora} of our model may also be possible. Personalization is an avenue we would like to explore formally in future work. For now, our code supports training by users, and our model dimensions were chosen carefully to enable this on consumer hardware. 
%Examples of maximal input and output sequence lengths of 1024 (the minimum amount we believe would be useful for multi-track MIDI infilling) can be trained on video cards with 6 GB RAM, and examples of maximal input and output sequence lengths of 1650 can be trained with 12 GB RAM. 
A video card with 6 GB of RAM is sufficient to train our released model on examples with input and output lengths of 1024,
%(the minimum amount we believe would be useful for our infilling task), 
and 12 GB of RAM is sufficient to train on examples with input and output lengths of 1650.
%Users with more RAM can train on longer examples. 
While this can benefit composers who wish to use our system, there is also the risk that our models may be trained by users to impersonate the styles of others.

\section{Conclusion}

We have introduced Composer's Assistant, a
%cross-platform 
system for interactive human-computer composition in the REAPER digital audio workstation. Composer's Assistant performs multi-track MIDI infilling and comes with an easy-to-use interface. We have released our source code, a pretrained model, a finetuned model, and scripts for interacting with our finetuned model in REAPER. Our models were trained only on permissively-licensed MIDI files.

\section{Acknowledgment}

We thank the many contributors to our MIDI training set for this project. Contributor acknowledgments can be viewed at our website.\footnotetwo We also thank the IT department at Sam Houston State University for building and maintaining the computational server on which we trained our model.

\bibliography{CA_bib}
\if\arxiv1
\bibliographystyle{plain}
\fi

\newpage 

\appendix
\onecolumn

\section{Supplemental information: Mask patterns for finetuning}

Our mask patterns for finetuning are given Table 3. We select from these patterns for building finetuning examples according to the following weights: 
pattern 0: 4/18, pattern 1: 6/18, pattern 2: 1/18, pattern 3: 1/18, pattern 4: 1/18, pattern 5: 1/18, pattern 6: 4/18. 
%\begin{itemize}
	%\item pattern 0: 4/18
	%\item pattern 1: 6/18
	%\item pattern 2: 1/18
	%\item pattern 3: 1/18 
	%\item pattern 4: 1/18
	%\item pattern 5: 1/18
	%\item pattern 6: 4/18
%\end{itemize}

\begin{table}[h!]
\label{TableMaskPatterns}
\begin{center}
\begin{tabular}{|p{0.45\linewidth} | p{0.45\linewidth}|}
\hline
\begin{tabular}{cc|ccccccc} & & \multicolumn{7}{c}{measure}\\ &   & 0 & 1 & 2 & 3 & 4 & 5 & 6 \\\hline\multirow{6}{*}{\rotatebox[origin=c]{90}{track}} & 0 & x &   &   &   & x & x &   \\ & 1 & x & x & x & x &   &   & x \\ & 2 & x & x & x & x & x & x &   \\ & 3 &   & x &   &   &   &   & x \\ & 4 & x &   & x &   &   &   &   \\ & 5 &   &   & x &   &   &   & x \\\end{tabular} 
& 
\begin{tabular}{cc|ccccccc} & & \multicolumn{7}{c}{measure}\\ &   & 0 & 1 & 2 & 3 & 4 & 5 & 6 \\\hline\multirow{6}{*}{\rotatebox[origin=c]{90}{track}} & 0 & x & x & x & x & x & x & x \\ & 1 &   &   &   &   &   &   &   \\ & 2 &   &   &   &   &   &   &   \\ & 3 &   &   &   &   &   &   &   \\ & 4 & x & x & x & x & x & x & x \\ & 5 &   &   &   &   &   &   &   \\\end{tabular}\\
Mask pattern 0: Random tracks and measures. Each track-measure is masked with probability 0.5.
& Mask pattern 1: Random tracks, all measures.\\
\hline
\begin{tabular}{cc|ccccccc} & & \multicolumn{7}{c}{measure}\\ &   & 0 & 1 & 2 & 3 & 4 & 5 & 6 \\\hline\multirow{6}{*}{\rotatebox[origin=c]{90}{track}} & 0 &   & x &   &   &   &   &   \\ & 1 &   & x &   &   &   &   &   \\ & 2 &   & x &   &   &   &   &   \\ & 3 &   & x &   &   &   &   &   \\ & 4 &   & x &   &   &   &   &   \\ & 5 &   & x &   &   &   &   &   \\\end{tabular}
&
\begin{tabular}{cc|ccccccc} & & \multicolumn{7}{c}{measure}\\ &   & 0 & 1 & 2 & 3 & 4 & 5 & 6 \\\hline\multirow{6}{*}{\rotatebox[origin=c]{90}{track}} & 0 &   & x &   &   &   &   &   \\ & 1 &   &  &   &   &   &   &   \\ & 2 &   & x &   &   &   &   &   \\ & 3 &   &   &   &   &   &   &   \\ & 4 &   & x &   &   &   &   &   \\ & 5 &   & x &   &   &   &   &   \\\end{tabular}\\
Mask pattern 2: Random measure, all tracks.
& Mask pattern 3: Random measure, most tracks.\\
\hline
\begin{tabular}{cc|ccccccc} & & \multicolumn{7}{c}{measure}\\ &   & 0 & 1 & 2 & 3 & 4 & 5 & 6 \\\hline\multirow{6}{*}{\rotatebox[origin=c]{90}{track}} & 0 & x & x & x & x &   &   &   \\ & 1 & x & x & x & x &   &   &   \\ & 2 & x & x & x & x &   &   &   \\ & 3 & x & x & x & x &   &   &   \\ & 4 & x & x & x & x &   &   &   \\ & 5 & x & x & x & x &   &   &   \\\end{tabular}
&
\begin{tabular}{cc|ccccccc} & & \multicolumn{7}{c}{measure}\\ &   & 0 & 1 & 2 & 3 & 4 & 5 & 6 \\\hline\multirow{6}{*}{\rotatebox[origin=c]{90}{track}} & 0 &   &   & x & x & x & x &   \\ & 1 &   &   &   &   &   &   &   \\ & 2 &   &   & x & x & x & x &   \\ & 3 &   &   & x & x & x & x &   \\ & 4 &   &   & x & x & x & x &   \\ & 5 &   &   & x & x & x & x &   \\\end{tabular}\\
Mask pattern 4: 2, 3, or 4 consecutive measures, all tracks.
& Mask pattern 5: 2, 3, or 4 consecutive measures, most tracks.
\\
\hline
\begin{tabular}{cc|ccccccc} & & \multicolumn{7}{c}{measure}\\ &   & 0 & 1 & 2 & 3 & 4 & 5 & 6 \\\hline\multirow{6}{*}{\rotatebox[origin=c]{90}{track}} & 0 &   &   &   &   &   &   &   \\ & 1 &   & x  & x  & x  &   &   &   \\ & 2 & x & x &   & x & x & x & x \\ & 3 &   &   &   &   &   &   &   \\ & 4 & x & x &   &   & x & x & x \\ & 5 &   &   &   &   &   &   &   \\\end{tabular} & 
\\
Mask pattern 6: Random tracks selected. Each selected track gets 1 or 2 random spans of masked measures.
& \\
\hline
\end{tabular}
\end{center}
\caption{Our mask patterns for finetuning, with an example of each provided. In the examples, x's indicate masked track- measures. Note that our ``last-bar mask'' for model evaluation is a special case of mask pattern 2 in this table.}
\end{table}

\section{Supplemental information: Additional training information}

As in \cite{t511}, we use a dropout rate of 0 for pretraining and a dropout rate of 0.1 (10\%) for fine-tuning. Dropout is disabled for inference. We use the PyTorch \cite{pytorch} versions of the Hugging Face \cite{huggingface} implementations of T5 \cite{t5} and Adafactor \cite{adafactor} for training. We use \tt{scale\_parameter=True} and \tt{relative\_step=True} as options for Adafactor. We experimented with switching to a fixed learning rate of $10^{-3}$ for finetuning (and switching these parameters to \tt{False}), as was done in \cite{t5}, and also with switching to a fixed learning rate of $10^{-3}$ only for the last few epochs of finetuning, but found that we achieved better results in any given amount of training time by sticking with learning rate decay for the entirety of the finetuning process. We use gradient accumulation to achieve an effective batch size of 128 (or as close to 128 as possible) throughout pretraining and finetuning.

\section{Supplemental information: Using our model in REAPER}

\begin{figure}[ht]
\centerline{\framebox{
\includegraphics[width=\columnwidth, trim={0 80 80 70},clip]{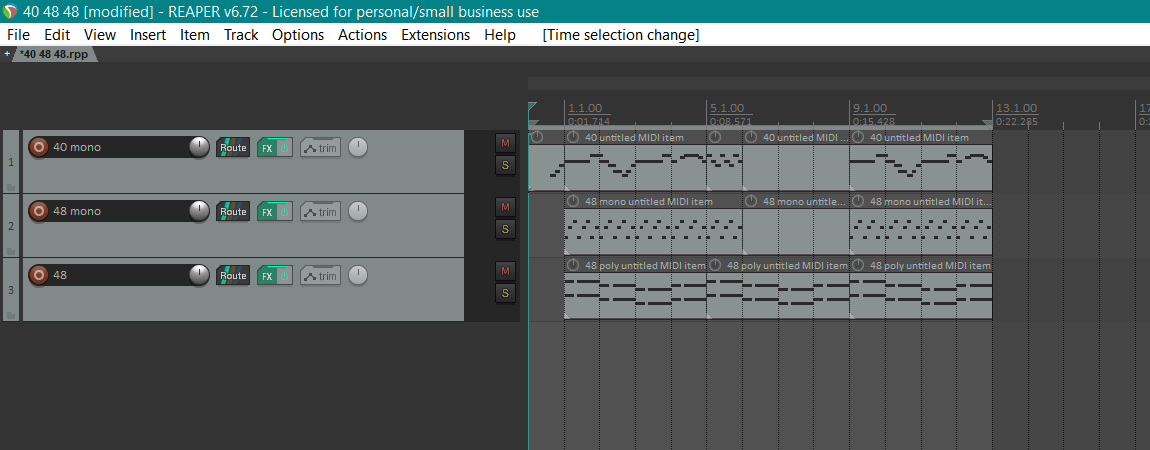}  % l b r t crop
}}
\centerline{\framebox{
\includegraphics[width=\columnwidth, trim={0 80 80 70},clip]{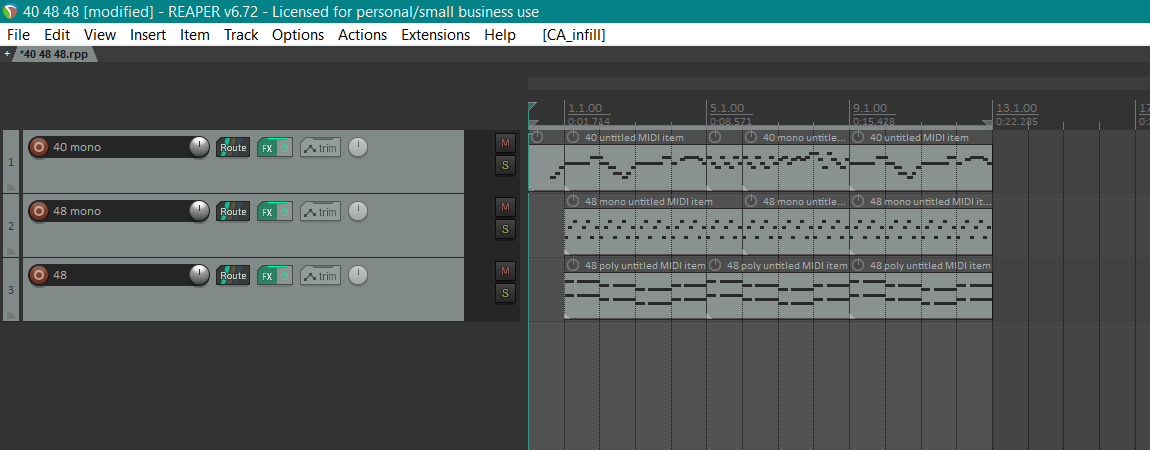}
}}
\caption{Using the model to connect two parts of a song, given a chord progression for the connection. The prompt is displayed above the output. Note that MIDI instruments 40 and 48 are ``violin'' and ``strings ensemble.''}
\label{FigREAPER2}
\end{figure}

\begin{figure}[ht]
\centerline{\framebox{
\includegraphics[width=\columnwidth, trim={0 80 80 70},clip]{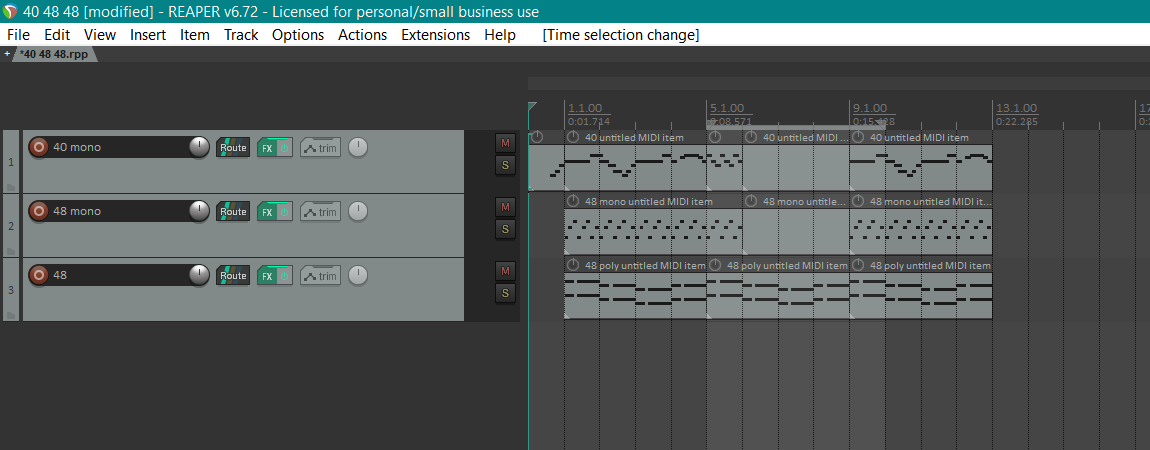}  % l b r t crop
}}

\centerline{\framebox{
\includegraphics[width=\columnwidth, trim={0 80 80 70},clip]{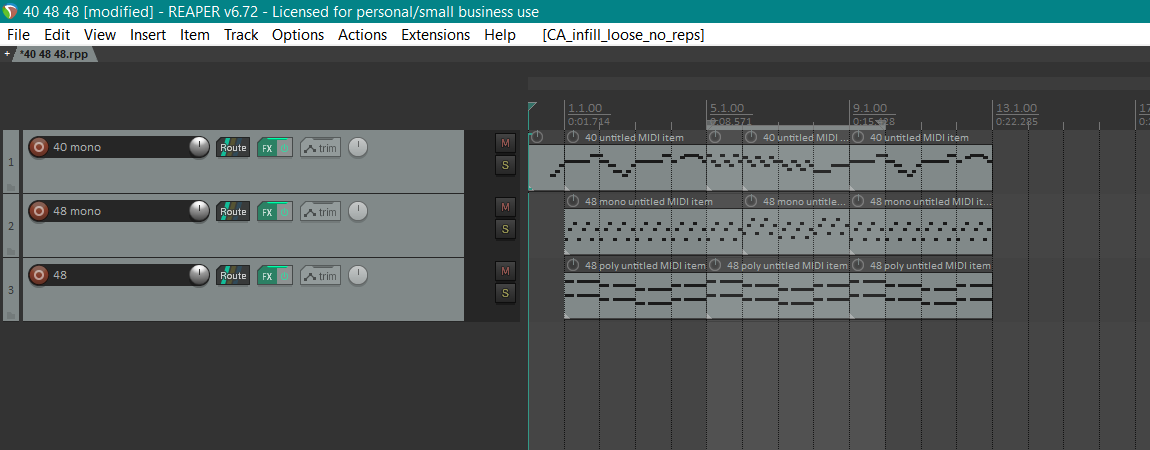}
}}
\caption{Using the model to connect two parts of a song, given a chord progression and part of the connection. The prompt is displayed above the output. The prompt contains only the selected measures. Sometimes it is useful to restrict which measures appear in the prompt, even if more measures could fit in memory, to prevent the model from writing parts that are too similar to parts that appear elsewhere in the song.}
\label{FigREAPER3}
\end{figure}

\begin{figure}
\centerline{\framebox{
\includegraphics[width=\columnwidth, trim={0 60 0 70},clip]{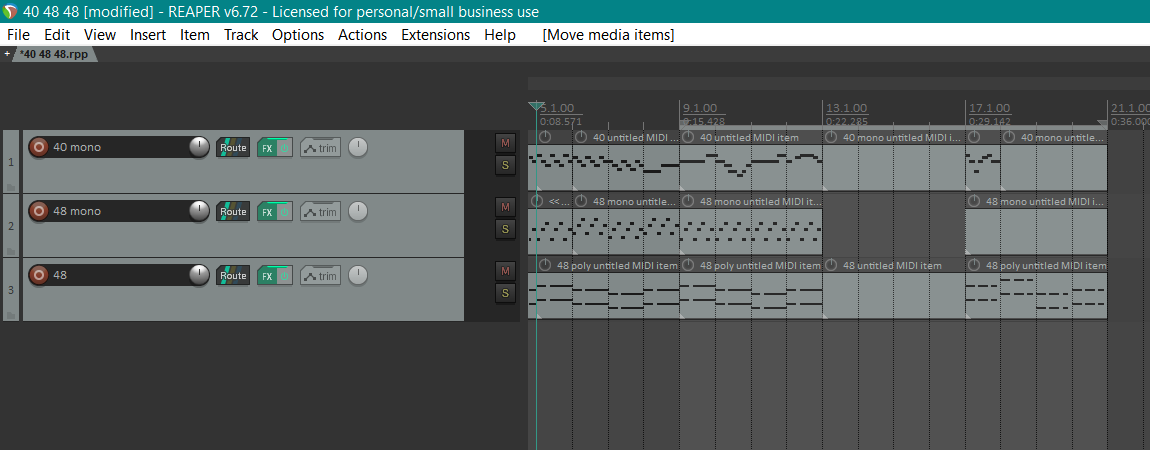}  % l b r t crop
}}

\centerline{\framebox{
\includegraphics[width=\columnwidth, trim={0 60 0 70},clip]{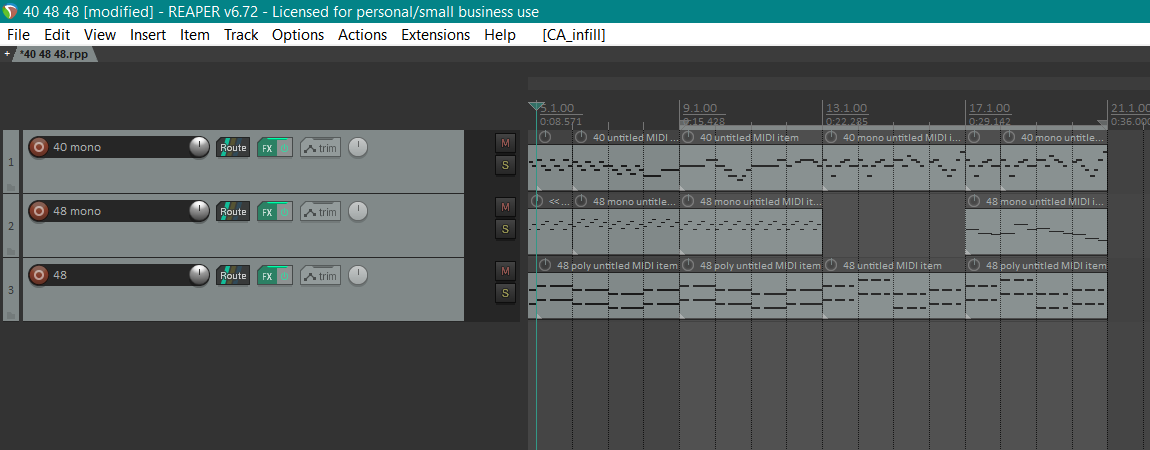}
}}
\caption{Using the model to continue a song, forcing a chord progression change and one measure of the melody later in the song. The prompt is displayed above the output. The model writes parts for the empty MIDI items, taking only the selected measures into account. Notice how the user selects the instrumentation for every measure of the prompt.}
\label{FigREAPER4}
\end{figure}

\begin{figure}
\centerline{\framebox{
\includegraphics[width=\columnwidth, trim={0 0 0 70},clip]{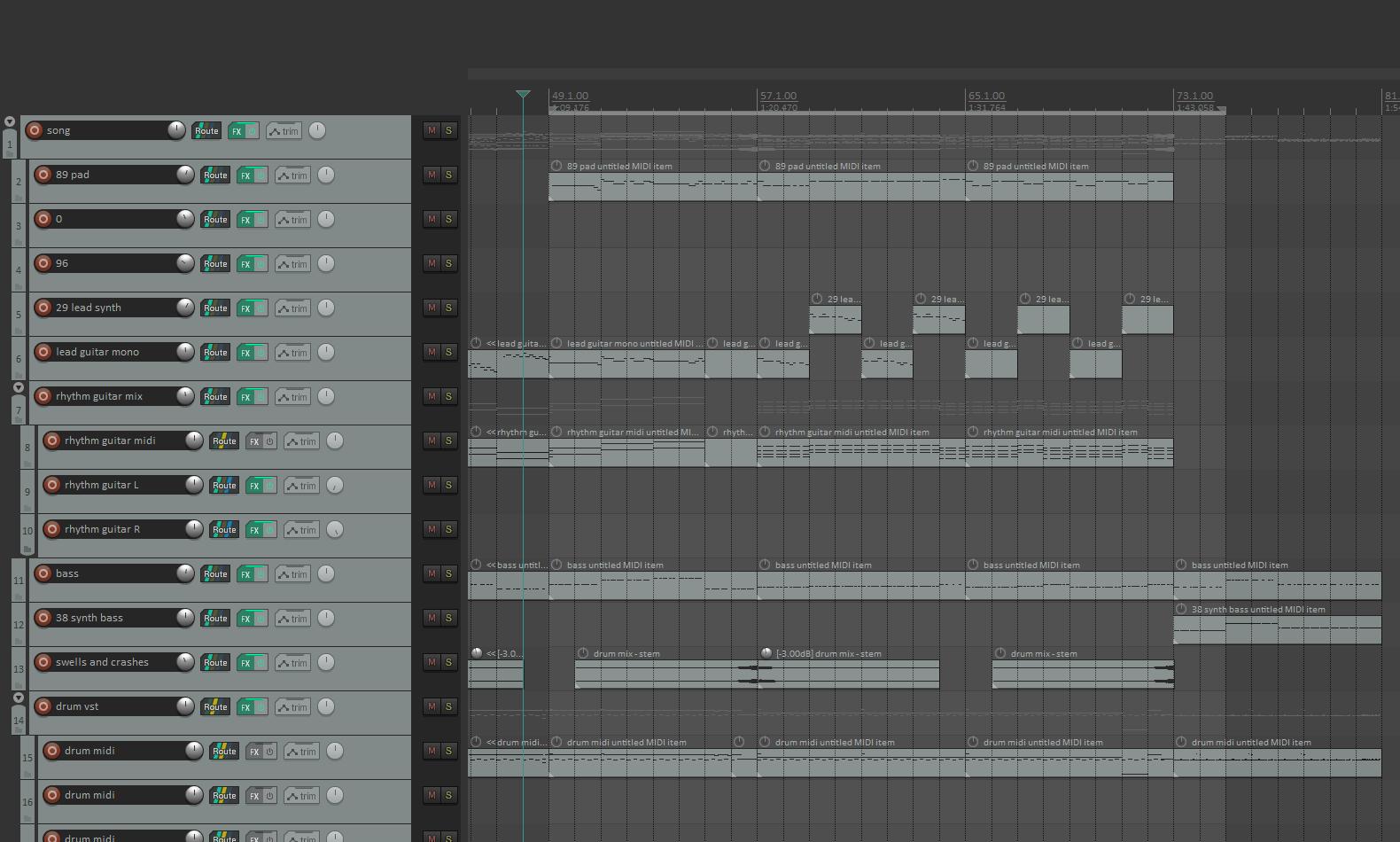}  % l b r t crop
}}
\caption{Using our model in large projects is straightforward. A prompt is displayed above. Our REAPER scripts ignore tracks without MIDI items, and infer MIDI instrument numbers from track names. Numbers at the beginning of a track name take precedence over words for determining the MIDI instrument of the track. See track 5: The track name ``29 lead synth'' includes in the model prompt a request to write for MIDI instrument 29 (overdriven guitar) for this track, even though the actual sound on the track is a lead synth sound. Since our model writes differently for each instrument, it is sometimes useful to lie to the model about what instrument is on a track.
Also note track 6: An explicit MIDI instrument number is not given in the track name (``lead guitar mono''), so MIDI instrument 30 (distortion guitar) is inferred from the ``guitar'' part of the name. Since ``mono" is included in the track name, the model prompt includes a request for a monophonic part for this track.}
\label{FigREAPERBIG}
\end{figure}

A REAPER user prompts our model by selecting a slice of measures in their project and then running one of our scripts. The script builds a model prompt from the selected measures, uses the model to evaluate the prompt, and writes the model's output back into the user's project. (If no measure selection is made, then the user's entire project is used to build the prompt.) 
Screenshots of us using our model to perform infilling can be found in Figures \ref{FigREAPER2}--\ref{FigREAPERBIG}. If a  user is unhappy with the output of one of our scripts, all they have to do to try again is hit ``undo'' in REAPER and run the script again, and the model will try to write something different from its previous output(s) on that prompt. 
Users can keep as many or as few of the notes output from the model as they like simply by editing their REAPER project before asking the model to try again.

Despite including no explicit chord conditioning or rhythmic conditioning in our training, we have observed that our model's outputs can be conditioned on chords and, to a certain extent, rhythms, by including a temporary instrument track  containing the desired chord progression played with the desired rhythm in a prompt. We have observed that a temporary piano (MIDI instrument 0) or strings ensemble (MIDI instrument 48) track works well for this purpose.

\end{document}